\documentclass[prd,twocolumn,aps,amsmath,nofootinbib,superscriptaddress]
{revtex4}

\usepackage{graphicx}
\usepackage{amsfonts}
\usepackage[usenames]{color}
\usepackage{amsmath}

\def\simgt{\mathrel{\lower2.5pt\vbox{\lineskip=0pt\baselineskip=0pt
           \hbox{$>$}\hbox{$\sim$}}}}
\def\simlt{\mathrel{\lower2.5pt\vbox{\lineskip=0pt\baselineskip=0pt
           \hbox{$<$}\hbox{$\sim$}}}}

\newcommand{\bea}{\begin{eqnarray}}
\newcommand{\eea}{\end{eqnarray}}
\newcommand{\nn}{\nonumber}

\newcommand{\vev}[1]{ \left\langle {#1} \right\rangle }

\newcommand{\comment}[1]{}

\begin{document}


\preprint{UCB-PTH-07/11, LBNL-62797, CALT-68-2653, UT-07-18}

\title{Quark and Lepton Masses from Gaussian Landscapes}

\author{Lawrence J.~Hall}
\affiliation{Department of Physics and Lawrence Berkeley National 
Laboratory,University of California, Berkeley, CA 94720, USA}

\author{Michael P.~Salem}
\affiliation{California Institute of Technology, Pasadena, CA 91125, USA}

\author{Taizan Watari}
\affiliation{Department of Physics, University of Tokyo, Tokyo, 
113-0033, Japan} 

\begin{abstract}
The flavor structure of the Standard Model might arise from random 
selection on a landscape. We propose a class of simple models, ``Gaussian 
landscapes,'' where Yukawa couplings derive from overlap integrals of 
Gaussian wavefunctions on extra-dimensions. Statistics of vacua are 
generated by scanning the peak positions of these zero-modes, giving 
probability distributions for all flavor observables. Gaussian landscapes 
can broadly account for all observed flavor patterns with very few free 
parameters.  For example, the generation structure in the quark sector 
follows from the overlap integrals for both the up and down type Yukawas 
sharing the localized wavefunctions of the quark doublets and the Higgs 
boson.  Although Gaussian landscapes predict broad probability 
distributions, the flavor observables are correlated and we show that 
accounting for measured flavor parameters creates sharper distributions 
for future neutrino measurements. 
\end{abstract}

\maketitle


\noindent
{\bf 1. {\em Introduction:}}
The Standard Model (SM) has enjoyed remarkable success at explaining 
laboratory data.  Nevertheless, it requires 28 parameters to be set by 
hand, thus begging for a more fundamental description.  Most of these 
parameters appear in the flavor interactions
\bea
{\mathcal L}_{\rm flavor} &=& \lambda^u_{ij}\, \bar{u}_i\, q_j \, h 
+ \lambda^d_{ij} \, \bar{d}_i\, q_j \, h^\dagger 
+ \lambda^e_{ij} \, \bar{e}_i\, l_j \, h^\dagger \nn\\
&& +\, \lambda^\nu_{ij} \, \overline{\nu}_i\, l_j \, h +
\lambda^M_{ij} \, \overline{\nu}_i\, \overline{\nu}_j \, \phi \,,
\label{eq:Lflav}
\eea
where $q,l$ ($\bar{u}$, $\bar{d}$, $\bar{e}$, $\overline{\nu}$) are 
the left (right) handed quark and lepton fields, $h$ is the Higgs boson, 
and $\phi$ represents the additional scalar(s) responsible for giving 
Majorana masses to the right-handed neutrinos $\overline{\nu}$.  Motivated 
by the success of unified gauge symmetries in describing the SM gauge 
couplings, the conventional wisdom is that some flavor symmetry is behind 
the patterns seen in the 22 flavor observables stemming from 
Eq.~(\ref{eq:Lflav}).  Indeed, it is well known that mass hierarchies and 
small mixing angles can arise from small flavor-symmetry breaking parameters 
when different SM generations feel different levels of the flavor-symmetry 
breaking~\cite{FN}.  On the other hand, we have yet to discover any 
precise, compelling relations among the flavor observables that would 
confirm a fundamental symmetry principle.

The cosmological dark energy may be evidence for a huge landscape of vacua, 
with the observed value of the cosmological constant resulting from 
environmental selection for large scale structure~\cite{CC}.  The current 
understanding of string theory seems to support the existence of such a 
huge landscape of vacua.  For example, in the Type IIB string description 
flux compactification generates a statistical distribution of complex 
structure moduli~\cite{flux}, which then determines the statistics of 
Yukawa couplings.  Thus, some kind of statistical randomness may be 
involved in the Yukawa couplings and this randomness may explain the 
absence of relations among masses, mixing angles, and CP phases that would 
otherwise reflect some fundamental symmetry principle. This philosophy has 
been pursued in Refs.~\cite{HMW, DDR}.

Can a theory of flavor using sheer randomness explain the various 
qualitative patterns among the flavor observables?  We consider seven 
major features of flavor to be 
i) the hierarchical masses in the quark and charged lepton sectors, 
ii) the pairing structure (i.e. small mixing angles) of the quark sector, 
iii) the generation structure of the quark sector (that is, the electroweak
pairing between the two heaviest, middle and lightest quarks),  
iv) the absence of pairing structure (i.e. large mixing angles) in the 
lepton sector, 
v) the mixing angle $\theta_{13}$ of the lepton sector being not so 
large as the other lepton mixing angles, 
vi) the hierarchy among Yukawa eigenvalues being largest in the up-quark
sector, and 
vii) the CP phase in the quark sector being of order unity.
Assuming simple statistical behavior of Yukawa couplings, Ref.~\cite{DDR} 
describes features (i) and (ii), while Ref.~\cite{HMW} describes 
feature (iv).  

Although these results are encouraging, they have several shortcomings.  
What is often broadly referred to as the generation structure of the SM 
includes features (i)--(iii), and is more confounding in light of the
large-mixing neutrino oscillations.  Thus we consider it crucial to
obtain features (i)--(iv) all at once in a single, simple framework.  
Refs.~\cite{HMW} and~\cite{DDR} use different schemes to describe the 
charged fermion and neutrino sectors, and fail to account for feature 
(iii).  Furthermore, these models lack a compelling motivation in terms of a 
more fundamental theory.  We introduce ``Gaussian landscapes'' as models of 
subsets of the landscape expected from compactification of Heterotic string 
theory.  As in Ref.~\cite{AS}, mass hierarchies, feature (i), arise without 
flavor symmetries due to small overlap integrals involving Gaussian 
wavefunctions on extra dimensions.  Whereas in Ref.~\cite{AS} there is no
landscape and these wavefunctions are positioned by hand to agree with data,
we take the peak positions to scan randomly over the geometry of extra 
dimensions, with each possibility corresponding to a distinct vacuum of the 
landscape.  Treating our universe as a typical, random selection from such a 
landscape, we find that Gaussian landscapes can broadly account for each of 
the seven features of flavor described above.  Ref.~\cite{HSW} gives a more
extensive account of the material in this letter.  

\vspace{10pt}
\noindent
{\bf 2. {\em Gaussian Landscapes:}}
In the compactification of the Heterotic string theory, Yukawa couplings 
are calculated by overlap integration of zero-mode wavefunctions 
over a six-dimensional compactification manifold. When the overlap of 
wavefunctions happens to be small, the relevant Yukawa couplings are small. 
It is known that the wavefunctions are approximately Gaussian on a base
manifold $B$ when a six-dimensional manifold is a torus fibration and 
the size of torus is small relative to $B$~\cite{Witten}.  
Meanwhile, if the gauge field moduli vary from one part of the universe to 
another, then the corresponding zero-mode wavefunctions vary, and likewise 
the masses and mixing angles.  The statistics of gauge field 
moduli,\footnote{These are dual to the complex structure moduli of the Type 
IIB string theory (F-theory) in cases with ${\cal N} = 1$ supersymmetry.} 
therefore, generate the statistics of flavor observables.  Statistics 
generated in this way are basis independent.

Aiming to extract the essence of string theory compactification with 
respect to flavor, we introduce the Gaussian landscape as a simplified 
version of the landscape.  The Gaussian landscape posits that low-energy 
degrees of freedom have localized (Gaussian) zero-mode wavefunctions in 
some geometry of extra dimensions, which is an analogue for the base 
manifold $B$.  In this letter we first use $S^1$ as the geometry, and 
briefly discuss the impact of the choice of geometry later.  The 
wavefunctions 
\begin{equation}
\varphi^a_i(y ; y^a_i) \simeq {\mathcal A}_i\, 
\exp\left[ {-(1+ir)(y - y^a_i)^2/(2 d_a^2) } \right]\, 
\label{eq:Gaussian}
\end{equation}
are taken to be complex with a phase related to the universal parameter $r$.
Here $a = q, \bar{u}, \bar{d}, l, \bar{e}, \overline{\nu}, h, \phi$ labels 
particle species and $i$ the generation of fermions.  The factor 
${\mathcal A}_i$ is a normalization factor chosen so that  
\begin{equation}
 M_5 \int_{0}^{L} dy \, |\varphi^a_i(y;y^a_i)|^2 = 1\,,
\label{eq:normalization}
\end{equation}
where $M_5$ is the cut-off scale of the four-dimensional effective theory 
and $L$ is the circumference of $S^1$.  In fact, the wavefunction 
(\ref{eq:Gaussian}) is made periodic on $S^1$, while maintaining 
the normalization in (\ref{eq:normalization}).  Yet as long as the width 
$d_a$ is parametrically smaller than the circumference $L$, the wavefunction 
is almost Gaussian.   

In the Gaussian landscape, the up-type Yukawa matrix derives from the 
overlap integral
\begin{eqnarray}
\lambda^u_{ij} = g M_5 \int_{S^1} dy\,\varphi^{\bar{u}}_i
(y; y^{\bar{u}}_i)\,
          \varphi^{q}_j(y; y^q_j)\,\varphi^h(y; y^h)\,,
\label{eq:overlap}
\end{eqnarray}
with $\lambda_{ij}^{d,e,\nu,M}$ determined analogously.\footnote{Note 
that we use $\varphi^h(y; y^h)$ also for the overlap integrations for 
$\lambda^{d,e}_{ij}$, which allows for the analytical understanding 
described later.  Since the wavefunction of the Higgs boson is not scalar 
valued if it originates from a vector field in extra dimensions, the 
``wavefunction'' $\varphi^{h^*}$ that is expected to be used
for $\lambda^{d,e}$ does not have to be the complex conjugate of
$\varphi^h$.  We emphasize that using a universal value for $r$ is not 
more than one of the simplest ways to introduce complex phases into 
Gaussian landscapes.}  
In this letter the coupling $g$ is assumed to be universal, as could 
result from higher-dimensional gauge interactions.
We assume very small neutrino masses are due to the see-saw mechanism.  
After integrating out the right-handed neutrinos $\overline{\nu}$, 
this generates a low-energy left-handed 
Majorana neutrino mass matrix from the effective interaction
\bea
\left(C_{ij}/\vev{\phi}\right)\, l_i\,l_j\,h\,h\,, \quad {\rm where} \quad
C_{ij} = 
\left( \lambda_\nu^T \, \lambda_M^{-1} \, \lambda_\nu \right)_{ij} \,.
\label{eq:see-saw}
\eea
To form the landscape we assume that the center coordinates $y^a_i$ of all 
wavefunctions are scanned freely and independently of one another on $S^1$.  
All of the other parameters are set by hand.\footnote{By slicing a subset 
out of a possibly much larger landscape of vacua, we see that there exists 
a subset that is phenomenologically successful.  It is a separate question 
whether such a subset is highly weighted in the vacuum statistics of the 
entire string landscape or because of cosmological evolution and/or 
environmental selection.  We consider that practical progress can be made 
by splitting the full problem into simpler parts.}
This random scanning of center coordinates is motivated by some knowledge
of instanton moduli; the center coordinates of multi-instantons can be 
chosen freely, and zero modes tend to localize around the centers of 
these instantons.\footnote{This is certainly not a rigorous argument; 
indeed basis independence of the probability distributions is also 
lost here (see Ref.~\cite{HSW}).  It would be interesting if this 
assumption---the random scanning of center coordinates---were refined by 
studying flux compactification. }

\begin{figure*}[t]
\begin{center}
\begin{tabular}{ccccc}
\includegraphics[width=0.19\linewidth]{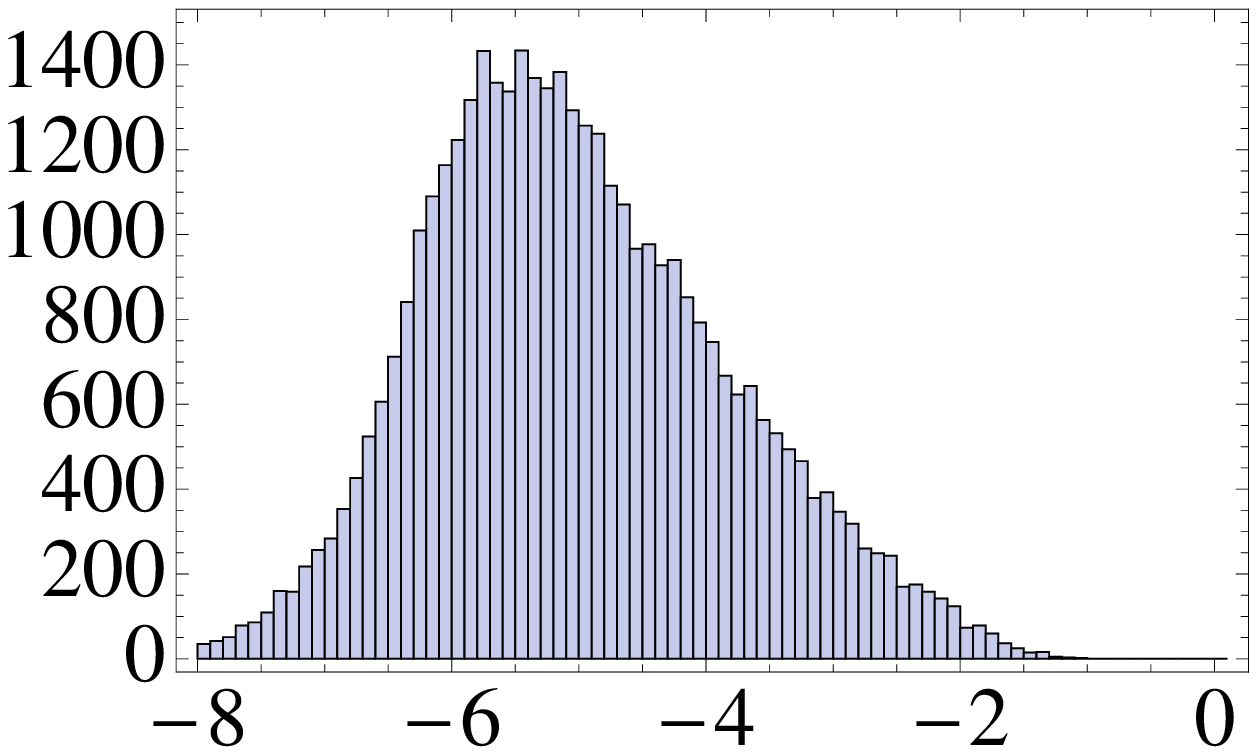} &
\includegraphics[width=0.19\linewidth]{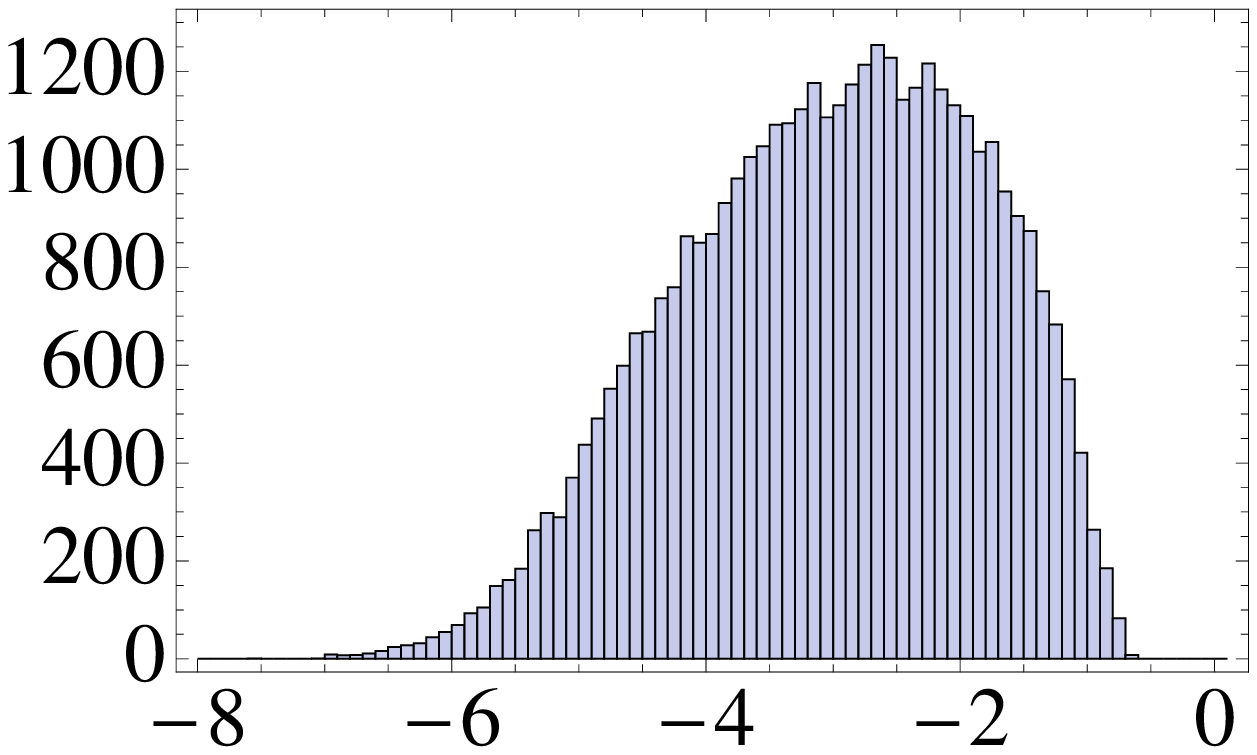} &
\includegraphics[width=0.19\linewidth]{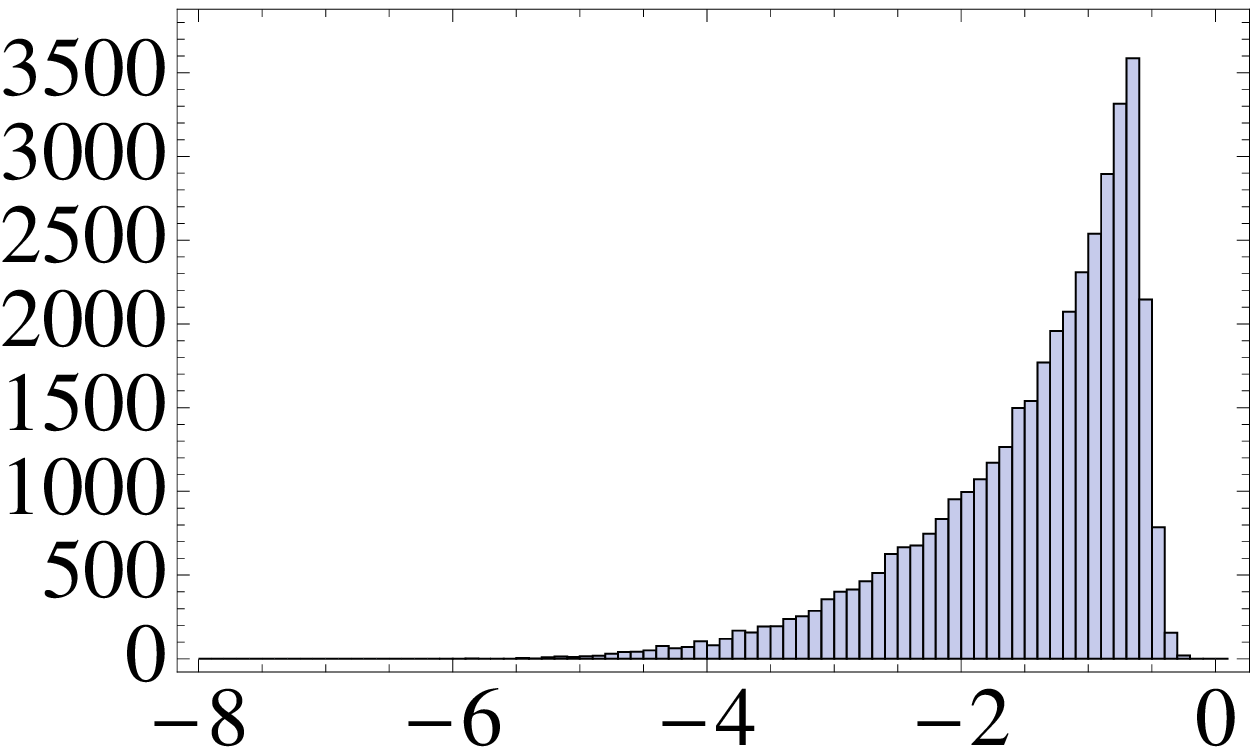} & 
\includegraphics[width=0.19\linewidth]{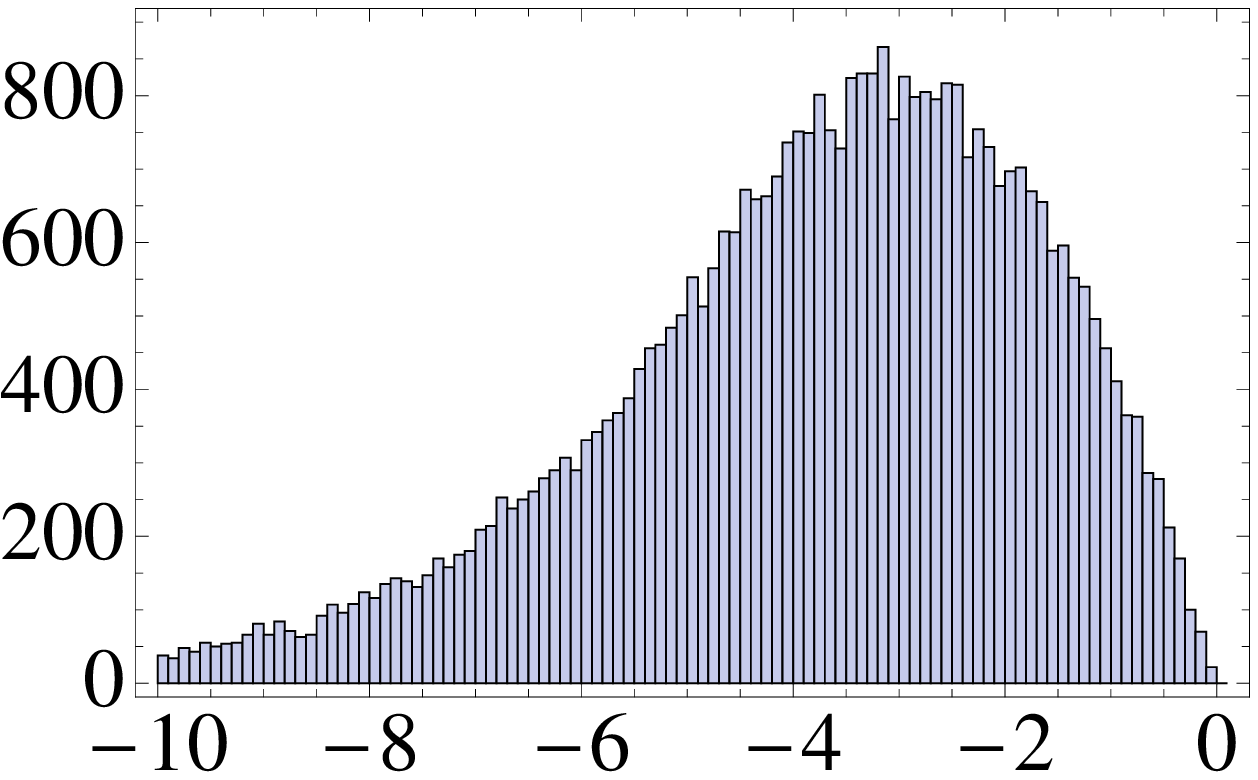} & 
\includegraphics[width=0.19\linewidth]{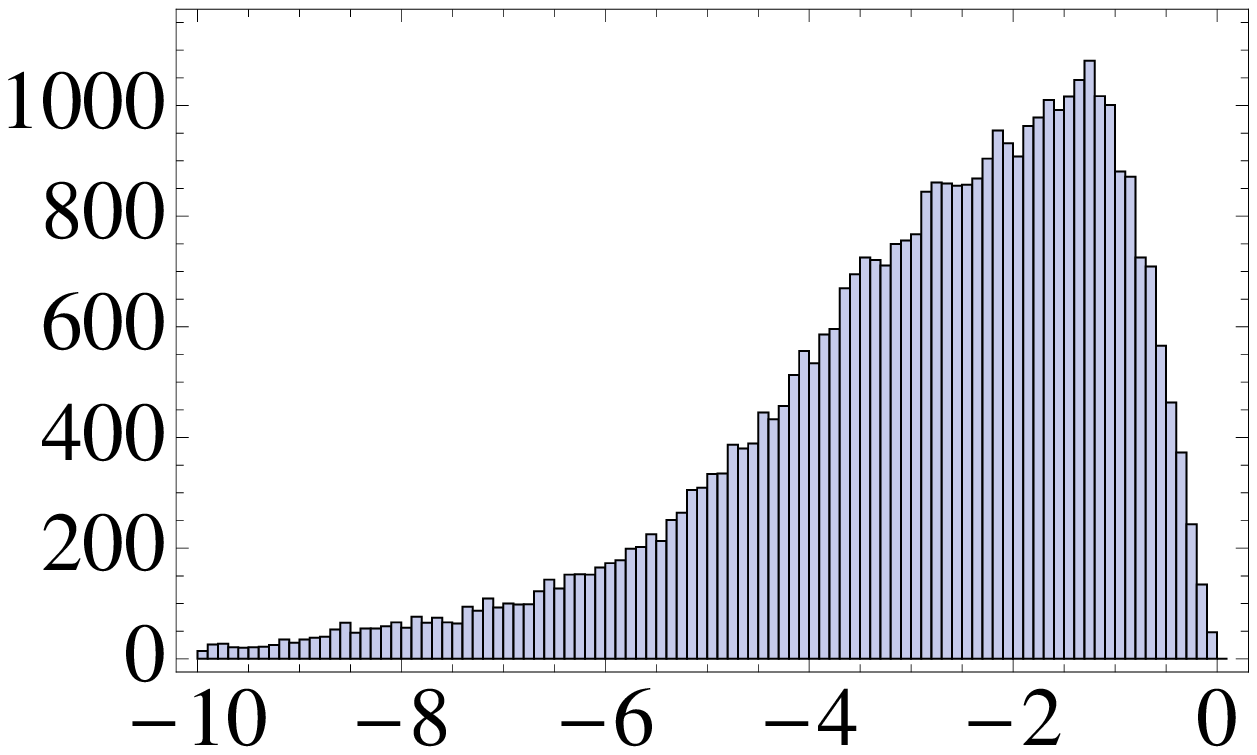} \\
$\log\lambda_{u}$ [$-5.5$] & 
$\log\lambda_{c}$ [$-2.9$] & 
$\log\lambda_{t}$ [$-0.3$] &  
$\log(m^\nu_1/m^\nu_2)$ &
$\log(m^\nu_2/m^\nu_3)$ \\
\includegraphics[width=0.19\linewidth]{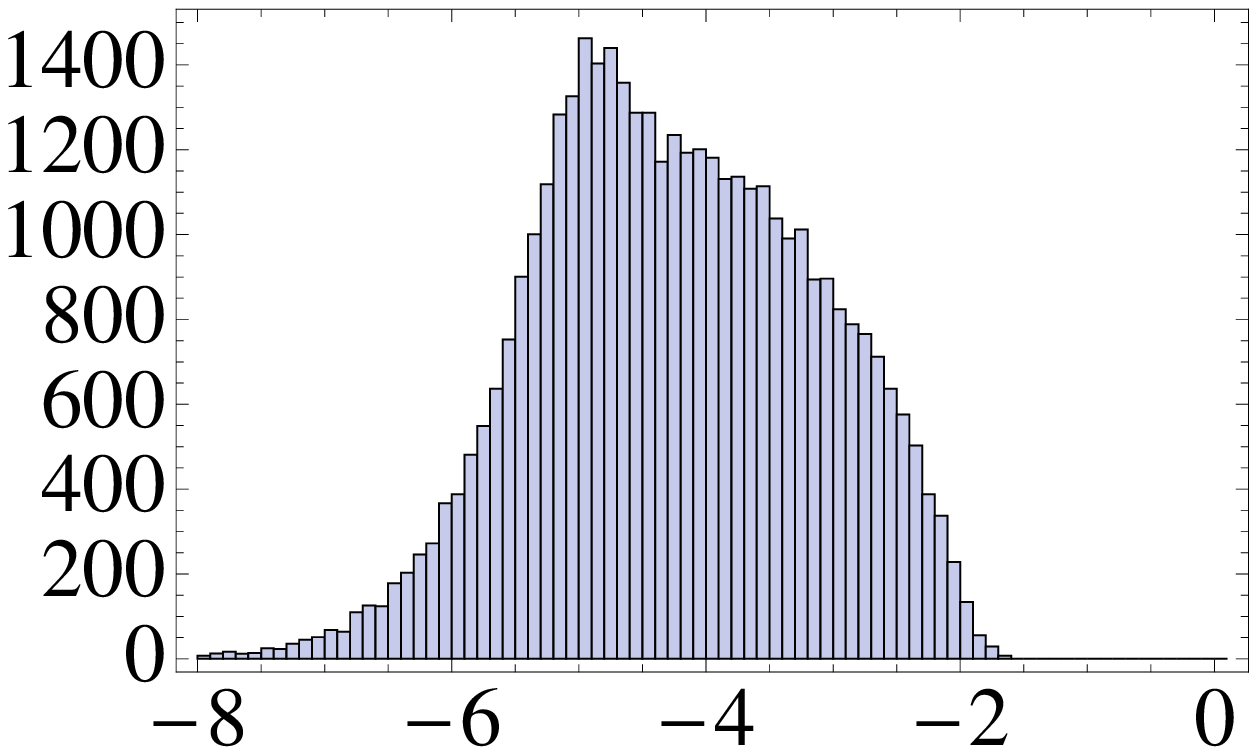} &
\includegraphics[width=0.19\linewidth]{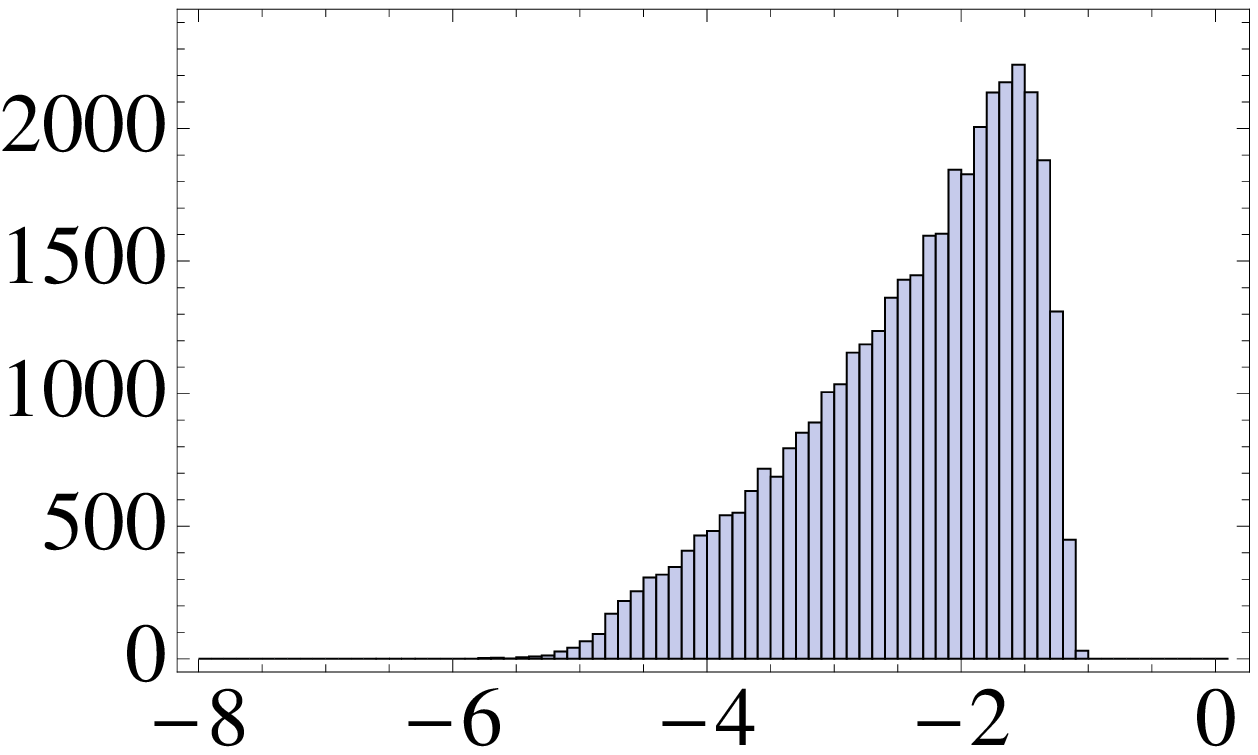} &
\includegraphics[width=0.19\linewidth]{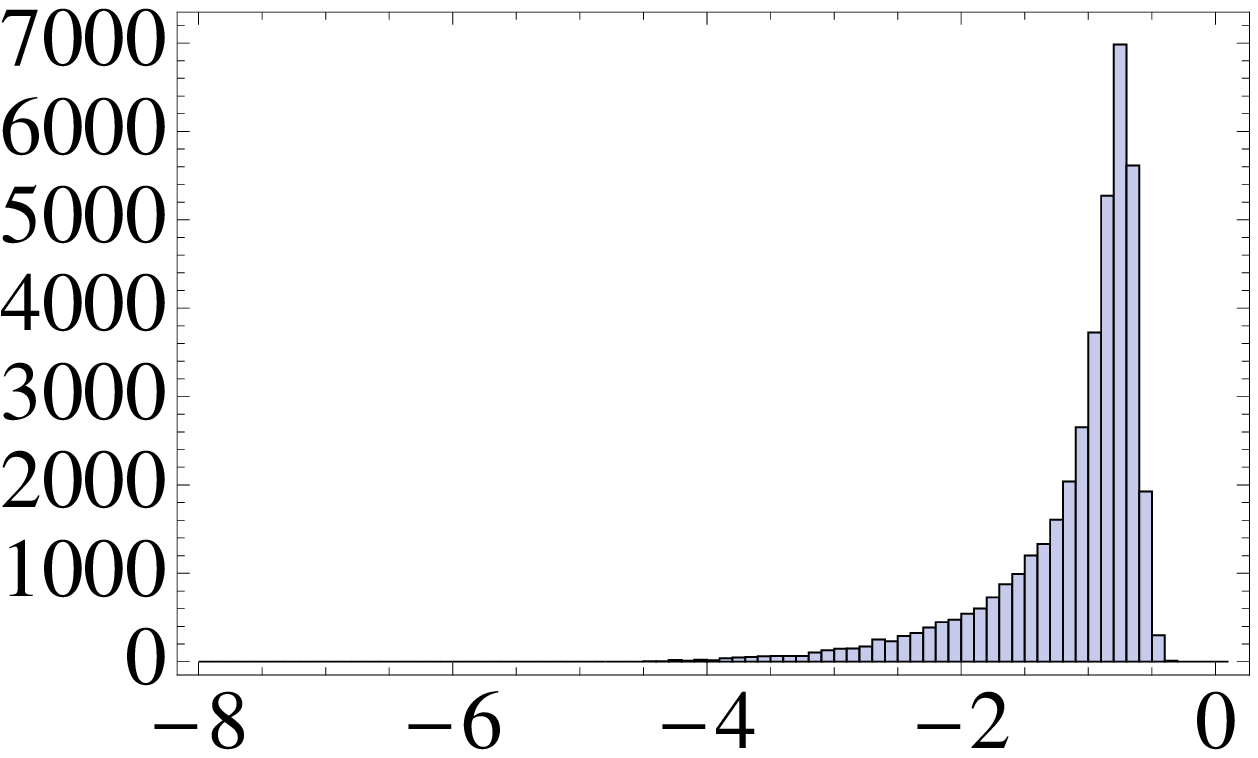} &
\includegraphics[width=0.19\linewidth]{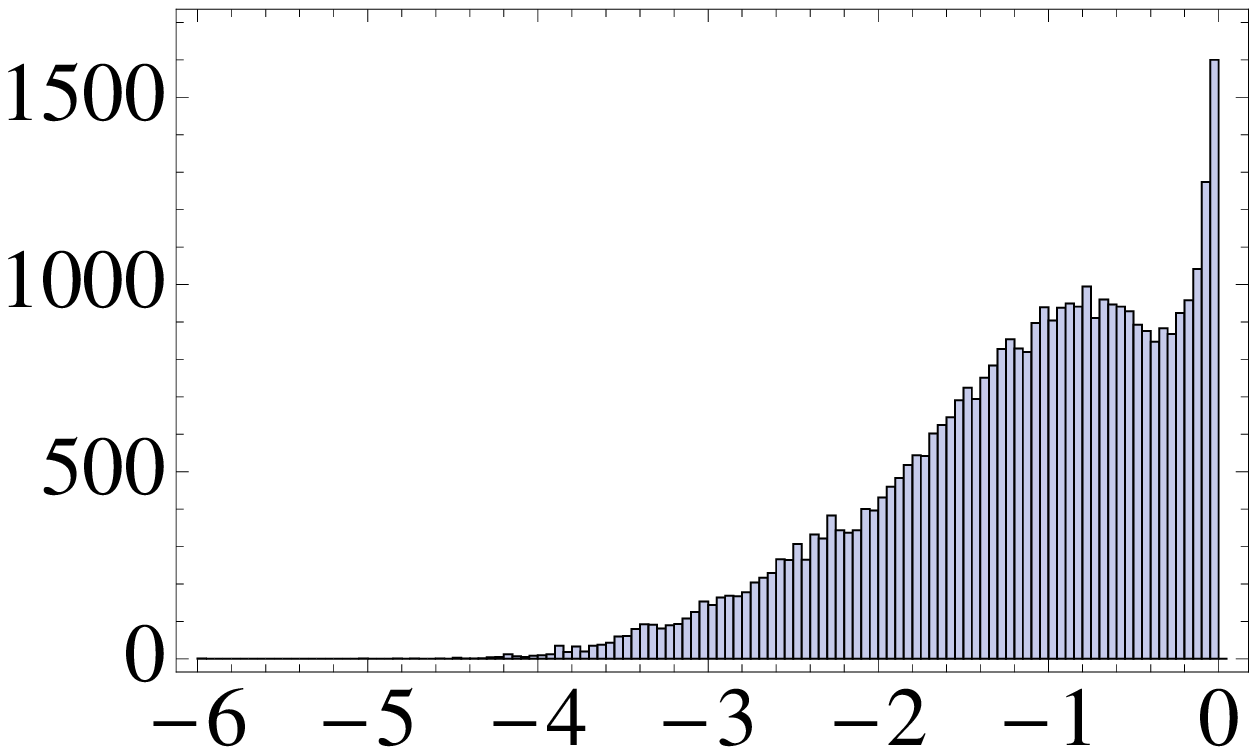} &
\includegraphics[width=0.19\linewidth]{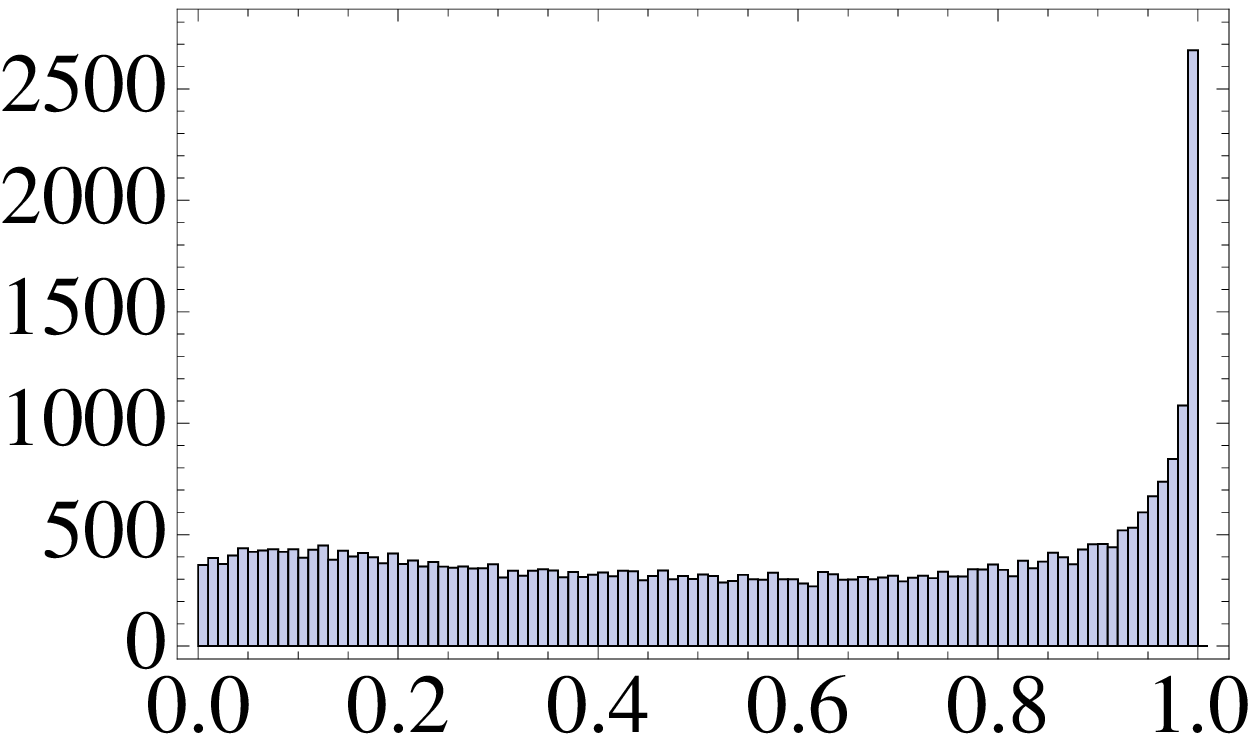} \\
$\log\lambda_{e,d}$ [$-5.5,-\!5.2$] & 
$\log\lambda_{\mu,s}$ [$-3.2,-\!3.9$] &
$\log\lambda_{\tau,b}$ [$-2.0,-\!2.2$] &
$\log(2\theta_{12}^{\rm CKM}/\pi)$ [$-0.8$] & 
$\sin (2\theta_{\odot})$ [0.91--0.93] \\
\includegraphics[width=0.19\linewidth]{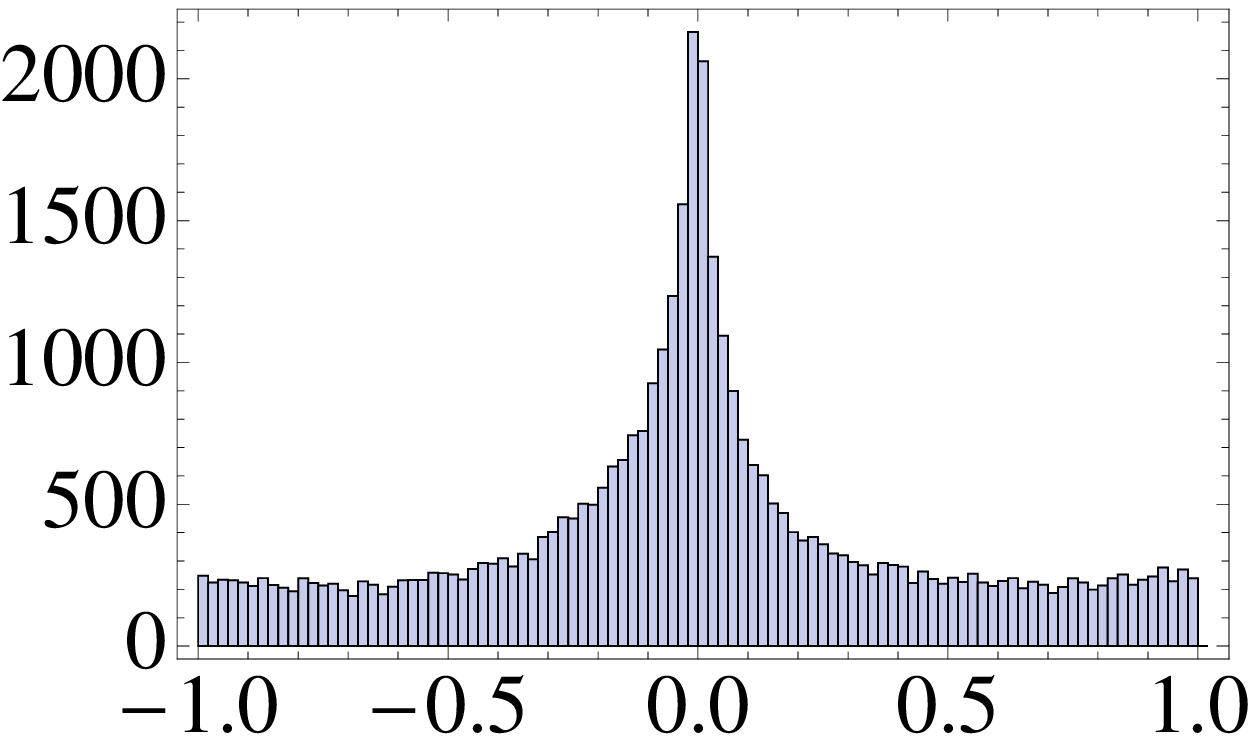} &
\includegraphics[width=0.19\linewidth]{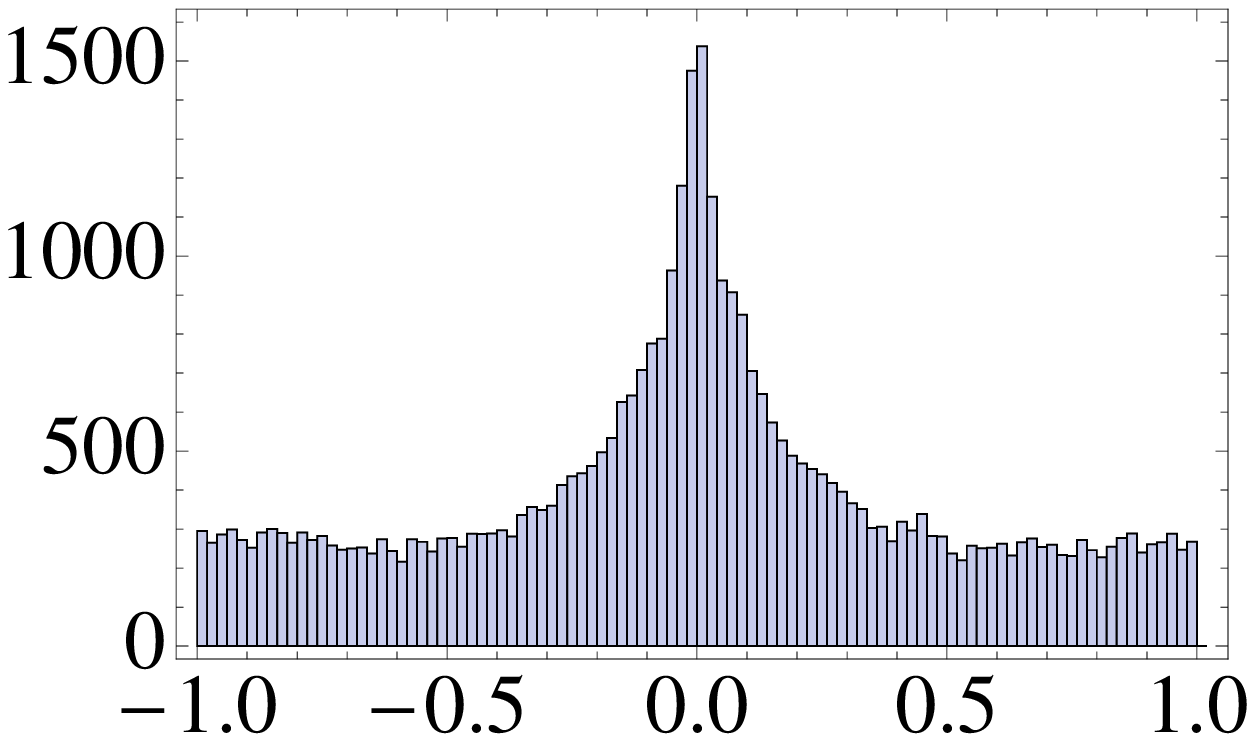} &
\includegraphics[width=0.19\linewidth]{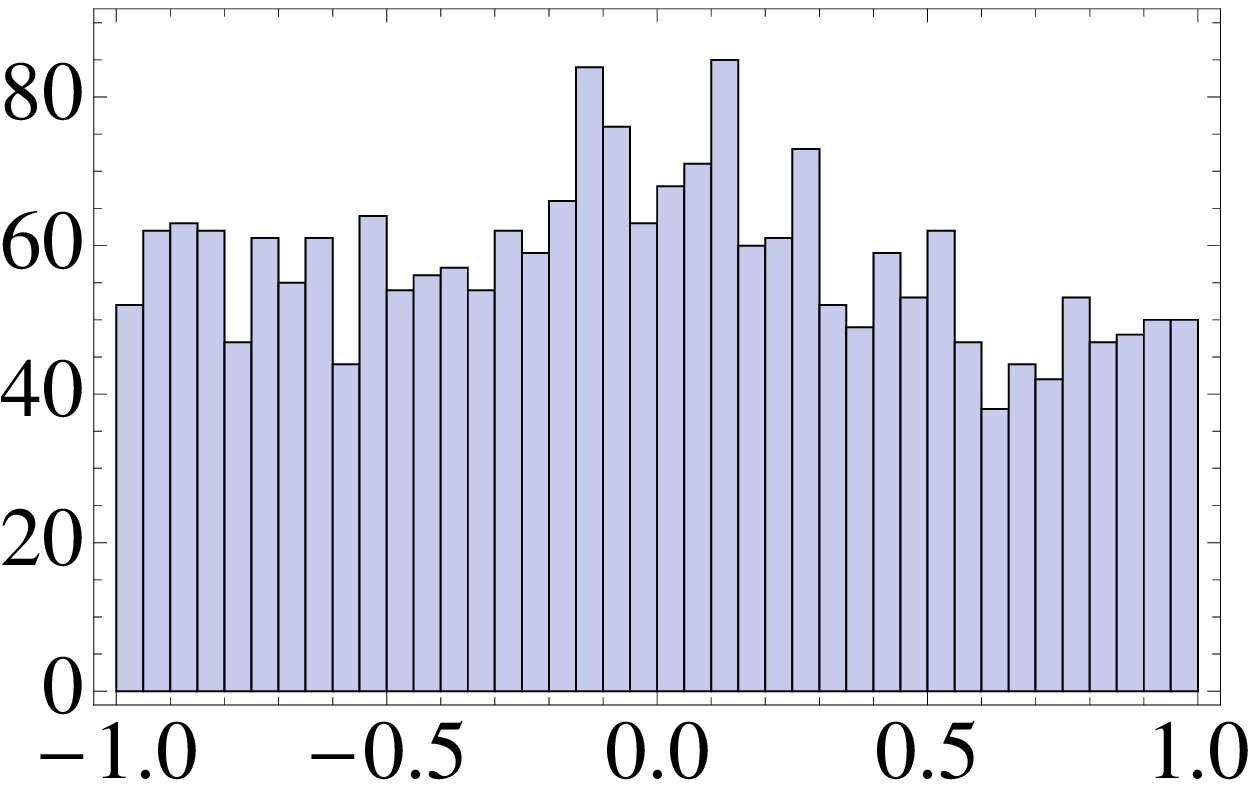} &
\includegraphics[width=0.19\linewidth]{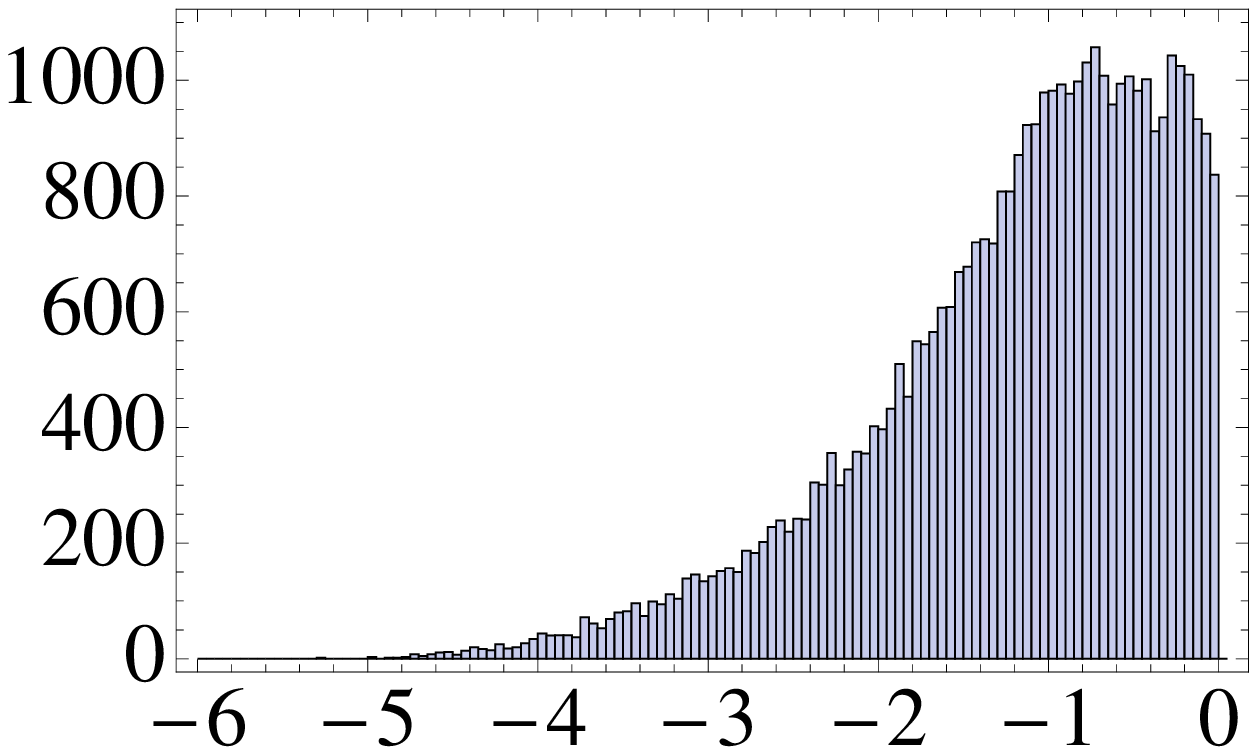} &
\includegraphics[width=0.19\linewidth]{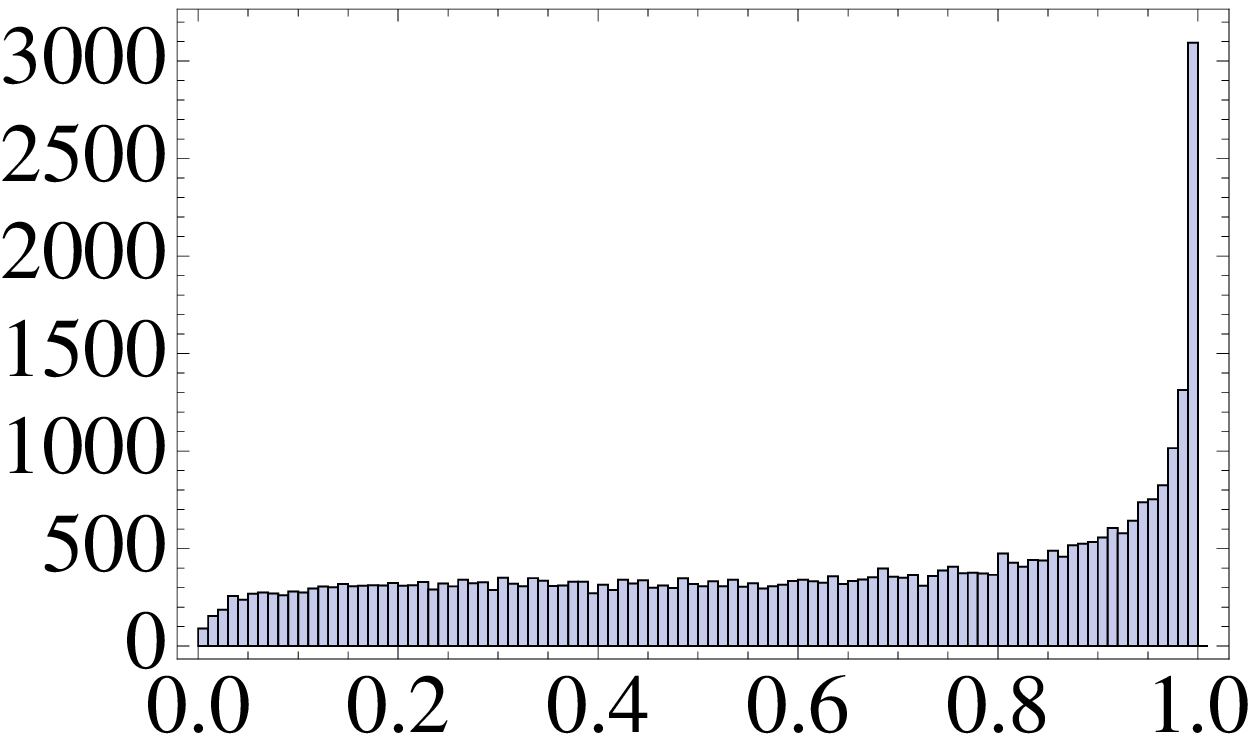} \\
$\delta_{\rm CKM}/\pi$ [0.29--0.34] &
$\delta_\nu/\pi$ &
$\delta_\nu/\pi$ (cuts $A$--$D$) &
$\log(2\theta_{23}^{\rm CKM}/\pi)$ [$-1.6$] &
$\sin(2\theta_{\rm atm})$ [$>$ 0.96] \\
\includegraphics[width=0.19\linewidth]{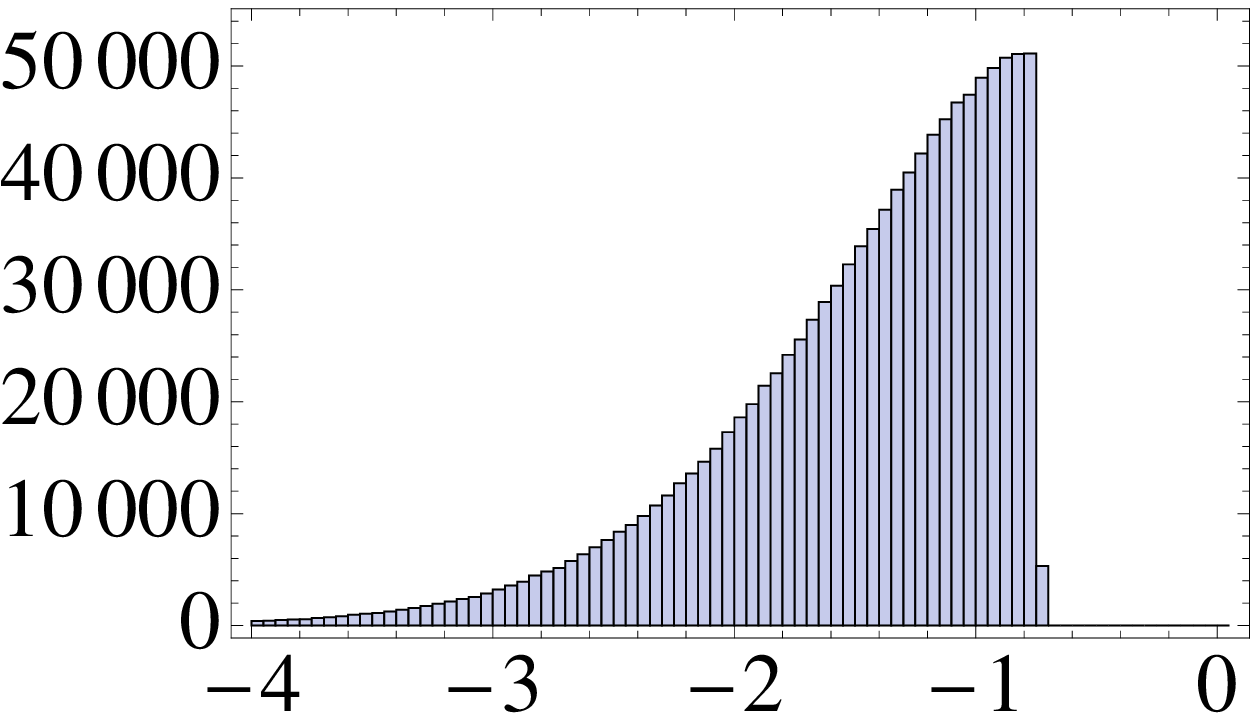} &
\includegraphics[width=0.19\linewidth]{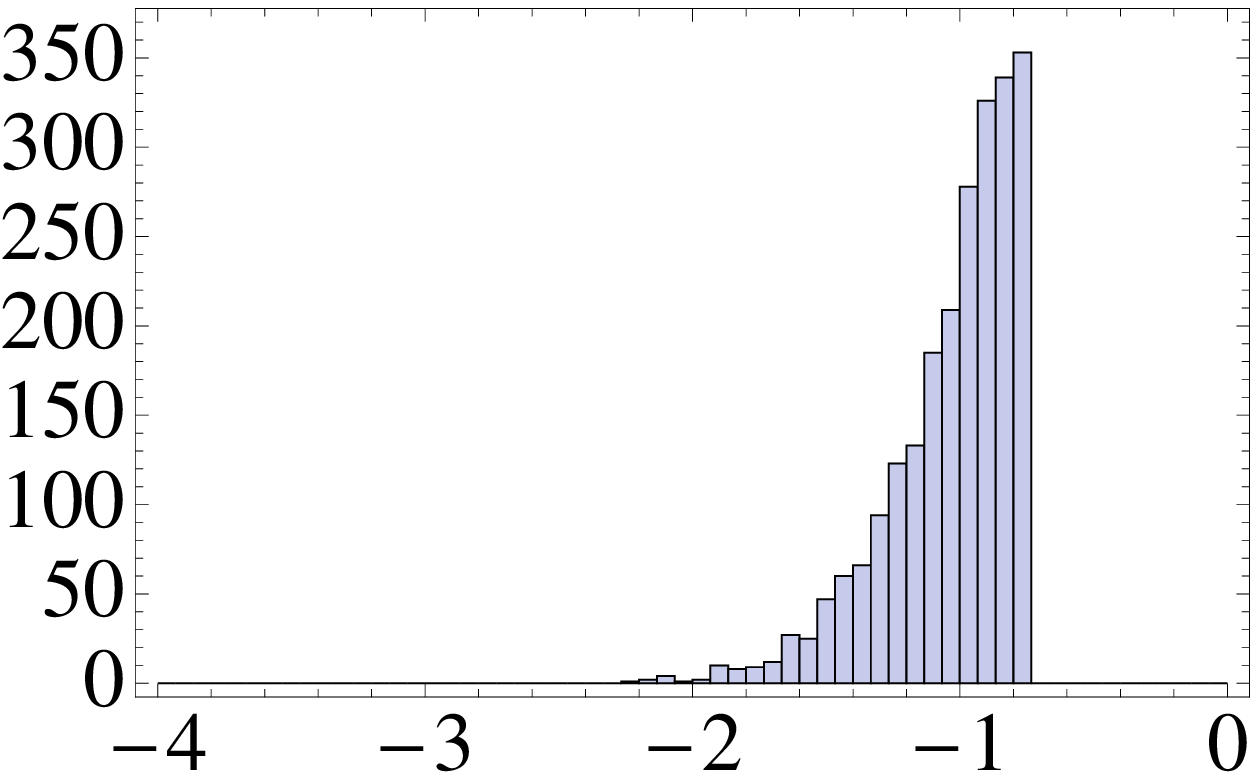} &
\includegraphics[width=0.19\linewidth]{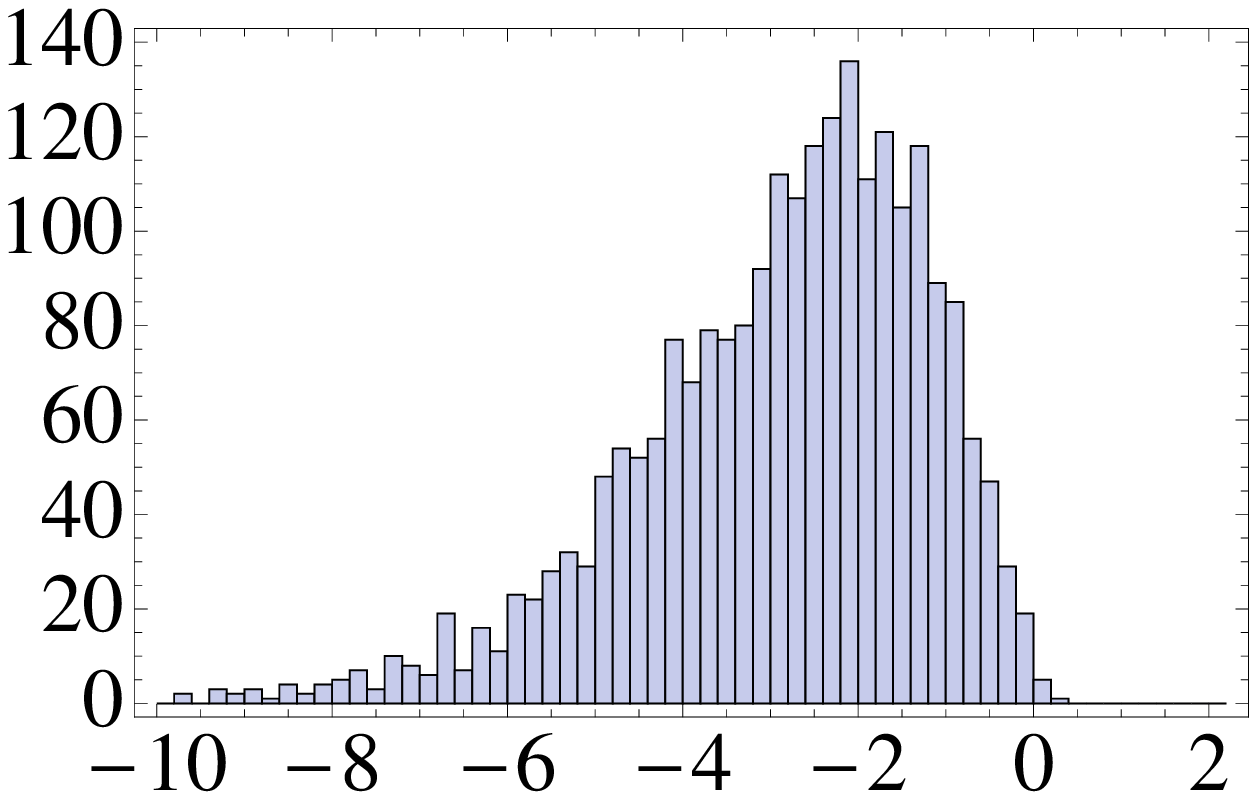} & 
\includegraphics[width=0.19\linewidth]{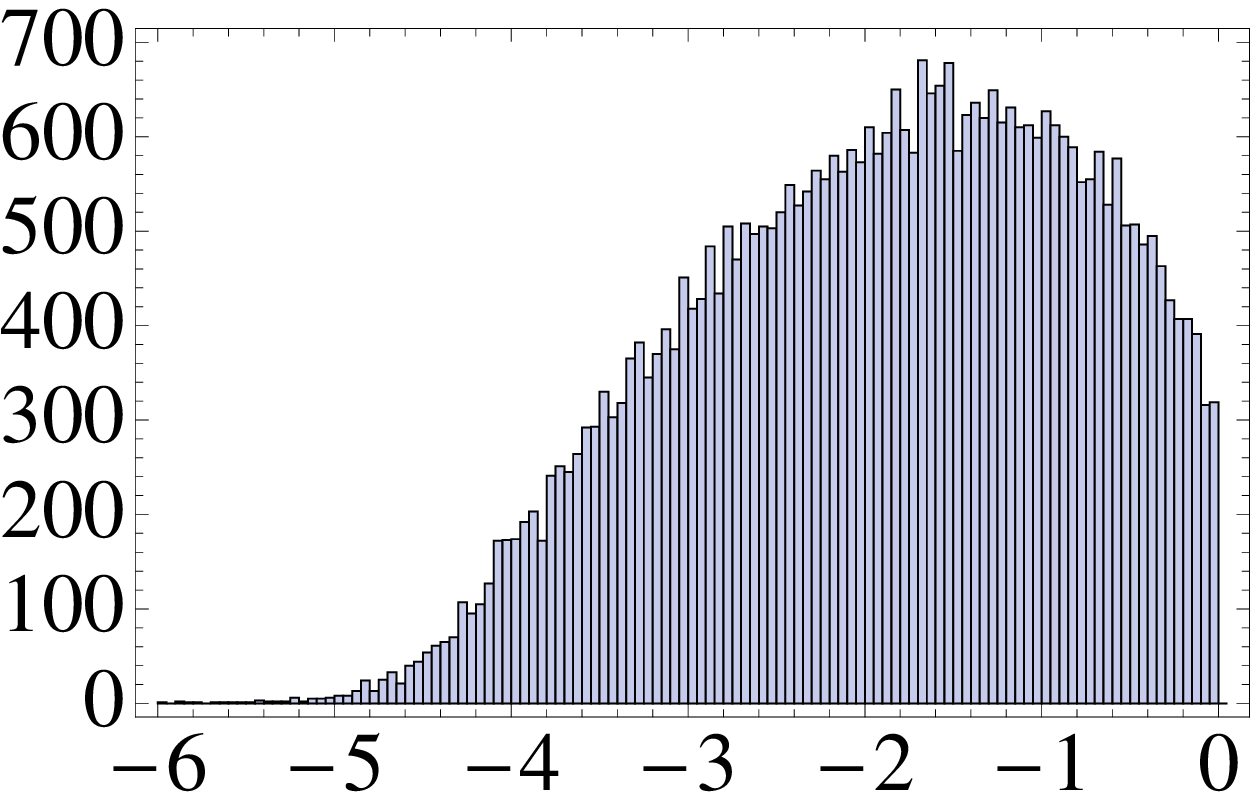} &
\includegraphics[width=0.19\linewidth]{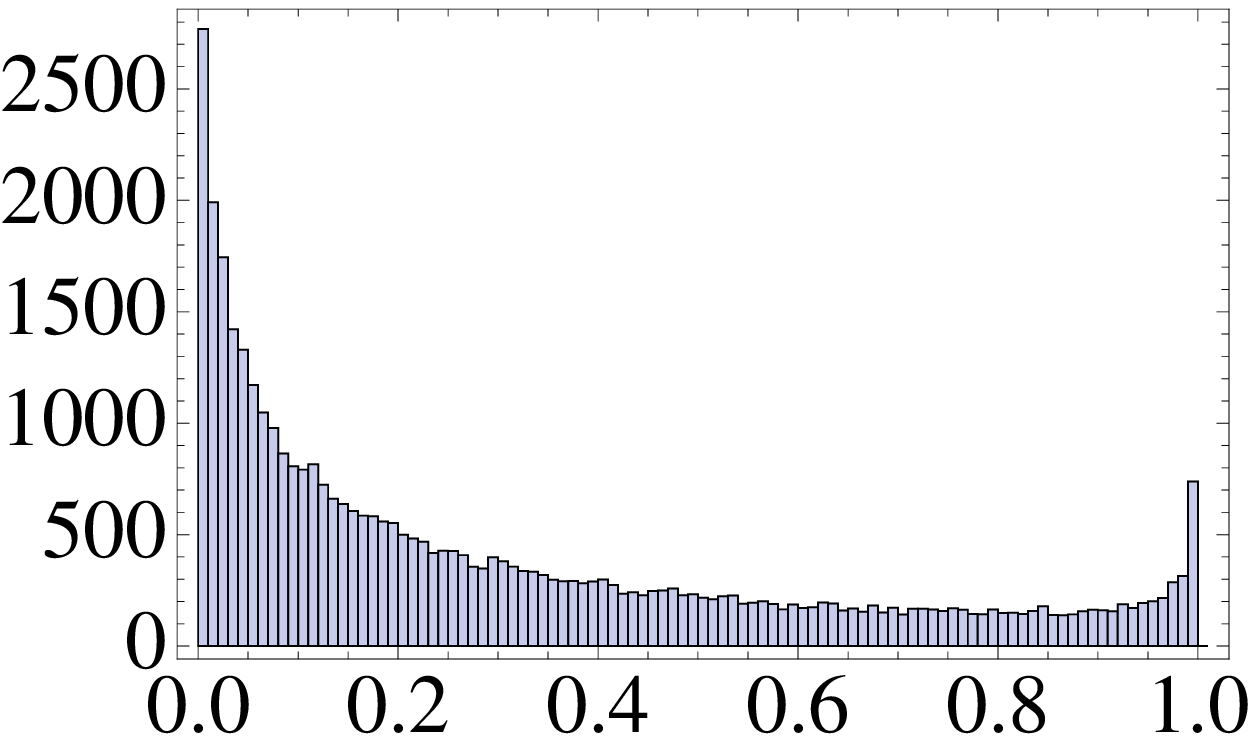} \\
$\log(\sin \theta_{13})$ (cut $D$) & 
$\log(\sin \theta_{13})$ (cuts $A$--$D$) & 
$\log|\Delta R|$ (cuts $A$--$D$) &
$\log(\sin\theta_{13}^{\rm CKM})$ [$-2.4$] &
$\sin (2\theta_{13})$ [$<$ 0.35] \\
\end{tabular}
\caption{\label{fig:complexS1} Distributions of observables in the $S^1$ 
Gaussian landscape. The width parameters are set at 
$d_h = d_{\bf 10} = d_{\bar{\nu}} = d_\phi = 0.08 L$ and 
$d_{\bar{\bf 5}} = 0.3 L$, and we use $r=3$ and $g=0.2$.
Numbers in brackets are the experimentally measured values (or limits), 
with the leading renormalization effects up to the Planck scale partially 
taken into account. All logarithms are base ten.}
\end{center}
\end{figure*}

Fig.~\ref{fig:complexS1} shows distributions of observables that follow 
from the Gaussian landscape on $S^1$.
Yukawa eigenvalues are hierarchical (i) and all three mixing angles in 
the quark sector are small (ii), if the wavefunctions of the Higgs boson 
and fields in the ${\bf 10}=(q,\bar{u}, \bar{e})$ multiplets are localized 
in the extra dimensions, i.e., $d_{\bf 10}, d_H \ll L$.  The overlap between 
a quark doublet $q_i$ and the Higgs boson introduces a correlation between 
the up-type and down-type Yukawa matrices, giving the observed electroweak 
pairing (i.e. no peaks at $\pi/2$ for the mixing angle distributions) in 
the quark sector (iii).  Thus, the localized wavefunction of the Higgs 
boson, as well as those of quark doublets, are the essence of what we 
observe as the generation structure in the quark sector.  We also find that 
the observed hierarchy 
$\theta_{13}^{\rm CKM} \ll \theta_{12}^{\rm CKM}, \theta_{23}^{\rm CKM}$ 
is typical.

The leptonic mixing angles are typically of order unity (iv),
when the width parameter $d_{\bar{\bf 5}}$ is set comparable to the size 
of the extra dimension, i.e. when the fields in $\bar{\bf 5}=(\bar{d},l)$  
do not have very localized wavefunctions.  
The angle $\theta_{13}$ tends to be smaller than the other two neutrino 
oscillation angles, agreeing very well with observation (v). The larger 
$d_{\bar{\bf 5}}$ width implies a milder hierarchy among the Yukawa 
eigenvalues in the charged lepton and down-quark sectors than in the 
up-quark sector (vi).  Although the distribution of CP phases depends on 
the details of how complex phases are introduced in the landscape, for 
the wavefunction (\ref{eq:Gaussian}) with $r=3$ there are continuous and 
flat components in the distributions of $\delta_{\rm CKM}$ and 
$\delta_\nu$ (vii).  We consider it unlikely that these flat components 
will disappear for minor modifications to the form of the wavefunctions.

Neutrino masses generated via the see-saw mechanism are typically 
very hierarchical in Gaussian landscapes, arising from both 
the overlap between $\overline{\nu}$ and $h$ in $\lambda^\nu$, and the 
overlap between $\overline{\nu}$ and $\phi$ in $\lambda^M$.  These two 
hierarchies are not correlated unless the Higgs boson and $\phi$ 
wavefunctions are correlated.  Therefore, in the statistical distribution 
of see-saw neutrino mass eigenvalues, these hierarchies add.  An inverted 
hierarchy is quite unlikely.  We note that complex phases play a crucial 
role in allowing for a sufficiently large value of 
$m_2^\nu/m_3^\nu$~\cite{HSW}.

The $S^1$ Gaussian landscape does not very well explain the large 
observed values of $\lambda_t/\lambda_{b,\tau}$.  Ref.~\cite{HSW} 
discusses how this problem can be addressed in 
Gaussian landscapes on geometries other than $S^1$.

\vspace{10pt}
\noindent
{\bf 3. {\em Analytic Approximation:}}
Most of the distribution functions in Fig.~\ref{fig:complexS1} can be 
understood with an analytic approximation.  Let us begin with the down-type 
Yukawa matrix.  The approximation is to consider $d_{\bf 10}, d_h\ll L$, 
with the various wavefunctions peaked not too far from each other, and 
$d_{\bar{\bf 5}}\approx {\cal O}(L)$.  Then
\begin{eqnarray}
\lambda^d_{ij}  \!\!&\propto&\!\! 
g\, \varphi_i^{\bar{d}}\left[y=(d_h^2 y^q_j + d_{\bf 10}^2 y^h)/
(d_{\bf 10}^2 + d_h^2) ;\, y^{\bar{d}}_i\right]
\epsilon^{q(d)}_j\, ; \label{eq:AFS-d} \nn\\
\epsilon^{q(d)}_j  \!\!&=&\!\! 
\exp \left[ - (1+r i)(y^q_j - y^h)^2/2(d_{\bf 10}^2 + d_h^2) \right].
\end{eqnarray}
The wavefunctions of the $\bar{d}_i$'s ($\subset \bar{\bf 5}$) are not
localized, yet not absolutely flat over $S^1$. 
With the random scanning of the center coordinates
$y^{\bar{d}}_i$, $y^{q}_i$ and $y^h$, the first factor $\phi_i^{\bar{d}}$ 
effectively yields random coefficients of order unity.  On the other hand, 
the second factor $\epsilon^{q(d)}_j$ can be exponentially small.
The quantity 
$\ln |\epsilon^{q(d)}| = - (\Delta y)^2/[2(d_{\bf 10}^2 + d_h^2)]$ 
can be as small as $-L^2/[8(d_{\bf 10}^2 + d_h^2)]$ in the $S^1$ 
Gaussian landscape, and its distribution function is determined 
only from the geometry of $S^1$: $dP/d|\Delta y|=2/L$, such that
\begin{equation}
 \frac{dP}{d |\!\ln\! |\epsilon^{q(d)}||}  = \sqrt{ 
    \frac{2(d_{\bf 10}^2 + d_h^2)}{L^2|\!\ln\! |\epsilon^{q(d)}| | } }
  = f(|\!\ln\! |\epsilon^{q(d)}||).
\label{eq:q(d)-FN}
\end{equation}
When a down-type Yukawa matrix is generated in the $S^1$ Gaussian 
landscape, the three $\ln\! |\epsilon^{q(d)}_j|$'s randomly follow a 
distribution function (\ref{eq:q(d)-FN}), and 
the largest, middle and smallest among them determine the order of 
magnitude of $\ln (\lambda_b/g)$, $\ln (\lambda_s/g)$ and 
$\ln (\lambda_d/g)$, respectively. Thus 
\begin{equation}
 \frac{d^3 P}{d |\!\ln \lambda_b|\, d|\!\ln \lambda_s| \, d|\!\ln \lambda_d|}
 \approx 3! f(|\!\ln \lambda_b|) f(|\!\ln \lambda_s|) f(|\!\ln \lambda_d|),
\label{eq:three-q(d)}
\end{equation}
for $g=1$.  The distribution of each of $\ln \lambda_{b,s,d}$ is obtained 
by integrating the two other variables. The resulting analytic approximate 
distribution agrees very well with the results of the numerical 
simulation above. 

The up-type Yukawa matrix has the structure 
\begin{equation}
 \lambda^u_{ij} = g_{ij}\epsilon_i^{\bar{u}}\epsilon_j^{q(u)} \,,
\label{eq:AFS-u}
\end{equation} 
for the widths $d_q = d_{\bar{u}} = d_{\bf 10}$, 
with statistically neutral random coefficients $g_{ij}$ and 
``flavor suppression factors''
\begin{equation}
 \epsilon^{q(u)}_j = \exp \left[ - 
    \frac{1+r i}{2d_{\bf 10}^2}
    \frac{d_{\bf 10}^2 + d_h^2}{d_{\bf 10}^2 + 2 d_h^2}
    (y^q_j - y^h)^2 \right] .
\end{equation}
The $\epsilon^{\bar{u}}_i$'s are given by the same expression except 
that $y^q_j$ is replaced by $y^{\bar{u}}_i$.
These flavor suppression factors follow a distribution function similar 
to (\ref{eq:q(d)-FN}), and 
the distribution functions of the largest, middle and smallest 
suppression factors are obtained just as in (\ref{eq:three-q(d)}).
The distribution function of $\ln \lambda_t$ ($\ln \lambda_c$ and 
$\ln \lambda_u$) is obtained by convoluting the distribution functions 
of the largest (middle and smallest) $\ln\epsilon^{\bar{u}}$ and 
$\ln\epsilon^{q(u)}$.

Similarly, approximate distribution functions can also be derived 
for the mixing angles of the quark sector and the mass eigenvalues of the 
charged leptons and see-saw neutrinos.  From the the flavor structure of the 
Yukawa matrices (\ref{eq:AFS-u}) and (\ref{eq:AFS-d}), hierarchical quark 
masses and small mixing angles in the quark sector follow from Gaussian 
landscapes, just as in the conventional flavor symmetry approach.  However, 
in Gaussian landscapes the geometry of the extra dimensions 
($dP/d (\Delta y)^2$ in particular) automatically determines the 
``flavor symmetry charge'' assignments within a given representation.  
Note also that the flavor suppression factor 
$\epsilon^{q(u)}_j$ associated with $q_j$ in the up-type Yukawa matrix 
is not the same as $\epsilon^{q(d)}_j$ in the down-type Yukawa matrix.

\vspace{10pt}
\noindent
{\bf 4. {\em Geometry Dependence:}}
So far we have considered only the Gaussian landscape on $S^1$.  
Ultimately we would like to understand
Gaussian landscapes with more extra dimensions and on non-trivial 
geometries. For example, three-dimensional base manifolds $B$, along 
with a $T^3$ fiber, can be used as six-dimensional manifolds in string 
theory compactification.  Therefore $B$ of more than one dimension 
will be of practical interest.  Such a study is launched in Ref.~\cite{HSW}; 
here we summarize initial results.  

The flavor structures (i)--(vi) observed in nature are obtained
statistically in Gaussian landscapes for various geometries $B$, 
as long as Gaussian wavefunctions on $B$ are localized for ${\bf 10}$'s 
and the Higgs but not for $\bar{\bf 5}$'s.  The distribution functions of 
observables certainly depend on the geometry of $B$, but in the analytic 
approximation, they depend only on the volume distribution 
$dP/d(\Delta y)^2$ of the geometry. The distribution functions of the 
observables are obtained by multiplying and integrating the volume 
distribution functions many times. Because of these integrations, the 
details in the volume distribution function  $dP/d(\Delta y)^2$ are smeared 
and only averaged properties of the geometry of $B$ are reflected in 
the distribution of observables. Some observables are more sensitive 
to the geometry, some not. See Ref.~\cite{HSW} for more details.

\vspace{10pt}
\noindent
{\bf 5. {\em Conditional Probability:}}
We have so far discussed vacuum statistics, assigning equal weight to 
each vacuum in the landscape. Questions of real interest, however, 
are probability distributions of observables with various weight factors 
from cosmological evolution and environmental selection included. 
Furthermore, we are interested in the distributions of yet-to-be
measured observables in the subset of vacua that pass all known 
experimental constraints. We hope to measure  one mixing angle 
$\theta_{13}$ and one CP phase $\delta_\nu$ in future neutrino oscillation
experiments, and one mass parameter $m_{\beta\beta}$ in neutrinoless 
double beta decay experiments.  
The leading contribution to $m_{\beta\beta}$ is $U_{e2}^2 m_2^\nu$:
$m_{\beta\beta} = U_{e2}^2 m_2^\nu ( 1+ \Delta R)$, 
but a precise prediction can be made only if the distribution for 
$|\Delta R| \equiv |(U_{e1}/U_{e2})^2 (m^\nu_1/m^\nu_2)|$ is 
peaked at small values.

It is not practical to obtain precise estimates of the weight 
factors or to generate statistics large enough so that all of the 
experimental cuts can be imposed. Instead, we use the Gaussian 
landscape on $S^1$ and extract a subset that satisfy ``loose
experimental cuts'' 
A) $10^{-2} < \Delta m^2_{\odot}/\Delta m^2_{\rm atm} < 10^{-1}$, 
B) $\sin^2 (2\theta_\odot) > 0.7$, 
C) $\sin^2 (2\theta_{\rm atm}) > 0.8$ and, 
D) $\sin \theta_{13} < 0.18$, to get a feeling for how much the 
distributions of $\theta_{13}$, $\delta_\nu$ and $|\Delta R|$ are 
affected by the weights and/or cuts.  The results are displayed in
Fig.~\ref{fig:complexS1}.  We see that imposing these cuts makes a 
significant difference in the expectations for future experiments.  
In particular the distribution for $\theta_{13}$ becomes peaked 
near the experimental limit, and there is no longer a contribution 
to the $\delta_\nu$ distribution that is peaked around zero.  
Finally, the distribution for $\Delta R$ is greatly reduced for all 
values of $\Delta R$ above 0.1, sharpening the prediction for 
$m_{\beta\beta}$ on the $S^1$ landscape.

\vspace{10pt}
\noindent
{\bf 6. {\em Conclusion:}}
We have shown that many features of flavor can be broadly understood
from a simple Gaussian landscape using just five free parameters.  
Distributions for the 17 measured flavor parameters, as well as 
for neutrino parameters that may be measured in future experiments, 
are shown in Figure 1.  The distributions are broad and result from 
scanning the centers of each quark, lepton and Higgs wavefunction 
randomly and independently.  Even if flavor originates from a 
landscape of vacua, it may be that some flavor observables are affected 
by environmental selection, for example the first generation masses 
$m_{u,d,e}$ may be selected for certain atomic and nuclear properties.

\vspace{10pt}
\noindent
{\bf\em Acknowledgments:}
This work was supported in part by the NSF grant PHY-04-57315 (LJH),
the US DOE under contract No. DE-AC03-76SF00098 (LJH) and 
contract No. DE-FG03-92ER40701 (MPS, TW), and the Gordon and Betty 
Moore Foundation (TW).

\end{document}